\documentclass[preprint,showpacs,preprintnumbers,amsmath,amssymb]{revtex4}

\usepackage{amssymb}
\usepackage{amsmath}
\usepackage{graphicx}
\usepackage{epsfig}
\usepackage{subfigure}
\usepackage{amsfonts}
\usepackage{CJK}

\begin{document}

\title{Transferring entangled states of photonic cat-state qubits in circuit QED}

\author{Tong Liu$^{1}$}
\author{Zhen-Fei Zheng$^{2}$}
\author{Yu Zhang$^{3}$}
\author{Yu-Liang Fang$^{1}$}
\author{Chui-Ping Yang$^{1}$}
\email{yangcp@hznu.edu.cn}

\address{$^1$Quantum Information Research Center, Shangrao Normal University, Shangrao 334001, China}
\address{$^2$Key Laboratory of Quantum Information, University of Science and Technology of China, Heifei 230026, China}
\address{$^3$School of Physics, Nanjing University, Nanjing 210093, China}

\begin{abstract}
We propose a method for transferring quantum entangled states of two photonic
cat-state qubits (\textit{cqubits}) from two microwave cavities to the other two
microwave cavities. This proposal is realized by using four microwave cavities
coupled to a superconducting flux qutrit. Because of using four cavities with different
frequencies, the inter-cavity crosstalk is significantly reduced. Since only one coupler qutrit is
used, the circuit resources is minimized. The entanglement transfer is completed with a single-step
operation only, thus this proposal is quite simple. The third energy level of the coupler qutrit is not
populated during the state transfer, therefore decoherence from the higher energy level is
greatly suppressed. Our numerical simulations show that high-fidelity transfer of two-\textit{cqubit} entangled states from
two transmission line resonators to the other two transmission line resonators
is feasible with current circuit QED technology. This proposal is universal and can be applied to accomplish the same task
in a wide range of physical systems, such as four microwave or optical cavities, which are coupled to a natural or
artificial three-level atom.
\end{abstract}

 \maketitle
\date{\today}

\section{Introduction}

Circuit quantum electrodynamics (QED), consisting of superconducting (SC)
qubits and microwave cavities or resonators, has developed fast in the past
decade and has been considered as one of the most promising platforms for
quantum information processing (QIP) \cite{s1,s2,s3,s4,s5,s6,s7,s8,s9}. SC qubits are good information
carriers and units of quantum information processors due to controllability
of their level spacings and recent significant improvement of their
coherence times. It was theoretically predicted that the strong-coupling
limit is readily achieved with SC charge qubits \cite{s3} or flux qubits \cite{s10}, and
later the strong and ultrastrong coupling between a SC qubit and a microwave
resonator was experimentally demonstrated \cite{s11,s12} (hereafter, the terms
cavity and resonator are used interchangeably). Based on circuit QED, many
proposals have been presented for implementing quantum state transfer
between SC qubits \cite{s1,s13,s14,s15}, quantum logic gates\ of SC qubits \cite{s16,s17,s18,s19,s20,s21}, and
entanglement in SC qubits \cite{s22,s23,s24,s25,s26,s27,s28}. By using SC qubits, the experimental
demonstrations of single-qubit gates \cite{s29,s30}, two-qubit gates \cite{s31,s32},
three-qubit gates \cite{s33,s34}, 10-qubit entanglement \cite{s35}, 12-qubit entanglement
\cite{s36}, 18-qubit entanglement \cite{s37}, and 20-qubit Schr\"{o}dinger cat states
\cite{s37} have been reported. Moreover, quantum teleportation between two distant
SC qubits \cite{s38}, quantum state transfer in a SC qubit chain \cite{s39},
entanglement swapping in superconducting circuit \cite{s40}, and quantum walks in
a 12-qubit superconducting processor \cite{s41} have been realized in experiments.

On the other hand, a (loaded) quality factor $Q=10^{6}$ for a
one-dimensional (1D) microwave resonator \cite{s42,s43} and a (loaded) quality
factor $Q=3.5\times 10^{7}$ for a three-dimensional (3D) microwave resonator \cite{s44} have been reported experimentally.
Photons, hosted by a microwave
resonator or cavity with a high quality factor, have much longer lifetimes
than SC qubits \cite{s45}. Hence, a microwave resonator or cavity of a high
quality factor can behavior as a quantum data bus and be used as quantum
memory. In recent years, there is much interest in quantum state engineering
and QIP with microwave fields or photons. Based on circuit QED, a number of
proposals have been presented for creating Fock states \cite{s46}, coherent states \cite{s47},
 squeezed states \cite{s48}, macroscopic Schr\"{o}dinger-cat states \cite{s49,s50,s51,s52},
and entangled states of microwave photons \cite{s53,s54,s55,s56,s57,s58,s59,s60,s61}. In addition, based on
circuit QED, how to realize two-qubit or multi-qubit quantum gates with
microwave photons has been investigated in theory \cite{s62,s63,s64}. The experimental
preparations of a Fock state or a superposition of Fock states of photons \cite{s65,s66,s67}, photonic Schr\"{o}dinger cat
states \cite{s68}, and photonic NOON states \cite{s69} have been reported. The coherent transfer of single photons between
microwave cavities has been demonstrated in experiments \cite{s70}. Moreover,
quantum error correction and universal gate set on a binomial photonic
logical qubit \cite{s71} have been experimentally demonstrated.

The focus of this paper is on photonic cat-state qubits (\textit{cqubits}).
For a cqubit, the two logic states are usually represented by two orthogonal
cat states (i.e., superposition states of coherent states) of photons.
In recent years, QIP with cqubits has attracted much attention because
coherent states
are eigenstates of the photon annihilation operator and tolerant to single-photon
loss~\cite{17nc,s72} and the
lifespan of a cqubit can be greatly improved by quantum error
correction~\cite{s72}. Proposals for entangling cqubits in a GHZ state~\cite{s73} and
for implementing single-cqubit gates~\cite{s74,s75}, two-cqubit gates~\cite{s58,s76}, and
multi-target-cqubit controlled gates~\cite{s77} have been presented. Moreover, the
experimental demonstration of single-cqubit gates~\cite{s78} and double-cqubit
entangled Bell states~\cite{s79} has been reported. However, after in-depth search
of literature, we found that how to directly transfer quantum entangled
states of \textit{cqubits} between cavities has not been studied yet.

In the following, we will propose a method to transfer quantum entangled
states of two cqubits from two microwave cavities to the other two microwave
cavities via circuit QED. This proposal is realized by using a
superconducting flux qutrit to couple four microwave cavities [Fig.~\ref%
{fig1}(a)]. Throughout this paper, \textquotedblleft
qutrit\textquotedblright\ refers to a three-level quantum system. As shown
below, this proposal has the following advantages: (i) Because of using four
cavities with different frequencies, the inter-cavity crosstalk is greatly
reduced; (ii) Due to the use of only one coupler qutrit, the circuit
resources is minimized; (iii) The entanglement transfer is quite simple
because only a single-step operation is needed; (iv) Because the higher
energy level $\left\vert f\right\rangle $ of the coupler qutrit is not
populated in the transfer process, the decoherence from this higher energy
level is greatly inhibited; and (v) Neither measurement on the cavity state
nor measurement on the qutrit state is required. In addition, our numerical
simulations demonstrate that high-fidelity transfer of two-cqubit entangled
states from two transmission line resonators to the other two transmission
line resonators is feasible with current circuit QED technology. This
proposal is universal and can be applied to transfer two-cqubit entangled
states from two microwave or optical cavities to the other two cavities,
which are coupled to a natural or artificial three-level atom.

This paper is organized as follows. In Sec.~II, we explicitly show how to
transfer quantum entangled states of two cqubits from two cavities to the
other two cavities. In Sec.~III, we give a discussion on the experimental
feasibility of the proposal. A concluding summary is presented in Sec.~IV.

\section{Transfer of quantum entangled states of two cqubits}

\begin{figure}[tbp]
\includegraphics[bb=40 610 468 782, width=11.5 cm, clip]{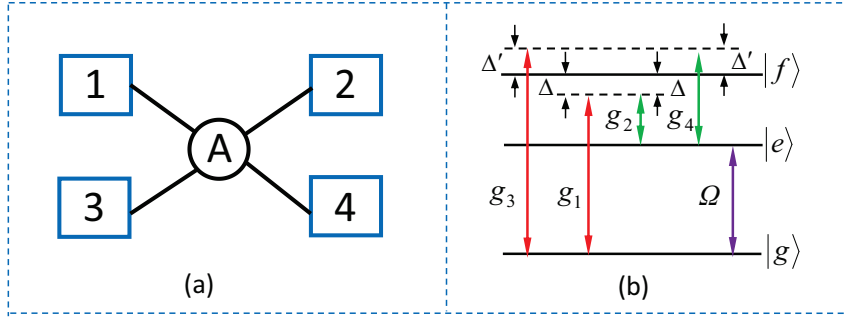} \vspace*{%
-0.08in}
\caption{(color online) (a) Diagram of a superconducting flux qutrit (a
circle $A$ at the center) and four microwave cavities. Each cavity here can
be a three-dimensional (3D) cavity or a one-dimensional (1D) cavity. For 3D
microwave cavities, the qutrit is inductively coupled to each cavity, by
placing the qutrit's partial loop area into each cavity. For 1D microwave
cavities, the qutrit is capacitively coupled to each cavity (see Fig. 2).
(b) Illustration of the qutrit-cavity dispersive interaction and the
qutrit-pulse resonant interaction.}
\label{fig1}
\end{figure}

Consider four microwave cavities ($1,2,3,4$) coupled to a superconducting
flux qutrit [Fig.~\ref{fig1}(a)]. The three levels of the coupler qutrit are
labeled as $\left\vert g\right\rangle ,$ $\left\vert e\right\rangle $ and $%
\left\vert f\right\rangle $ [Fig.~\ref{fig1}(b)]. Suppose that cavity $1$ ($%
3 $) is dispersively coupled to the $\left\vert g\right\rangle $ $%
\leftrightarrow $ $\left\vert f\right\rangle $ transition with coupling
strength $g_{1}$ ($g_{3}$) and detuning $\Delta $ ($\Delta ^{\prime }$),
cavity $2$ ($4$) is dispersively coupled to the $\left\vert e\right\rangle
\leftrightarrow \left\vert f\right\rangle $ transition with coupling
strength $g_{2}$ ($g_{4}$) and detuning $\Delta $ ($\Delta ^{\prime }$), and
each cavity is highly detuned (decoupled) from other energy level
transitions [Fig.~1(b)]. Here, $\Delta =\omega _{fg}-\omega _{1}=\omega
_{fe}-\omega _{2}>0$ and $\Delta ^{\prime }=\omega _{fg}-\omega _{3}=\omega
_{fe}-\omega _{4}<0,$ $\omega_{fg}$ ($\omega_{fe}$)
is the $|g\rangle\leftrightarrow|f\rangle $
($|e\rangle\leftrightarrow|f\rangle $) transition frequency of the flux qutrit,
and $\omega _{j}$ is the frequency of cavity $j$ ($%
j=1,2,3,4$). In addition, apply a microwave pulse to the qutrit. The pulse
is on resonance with the $\left\vert g\right\rangle $ $\leftrightarrow $ $%
\left\vert e\right\rangle $ transition but highly detuned (decoupled) from
other energy level transitions [Fig.~\ref{fig1}(b)]. These conditions can be
met by prior adjustment of the qutrit's level spacings or the cavity
frequencies. Note that the level spacings of a superconducting qutrit can be
rapidly (within $1-3$ ns) adjusted~\cite{s80} and the frequency of a microwave
cavity can be fast tuned with a few nanoseconds~\cite{s81}.

Under the above considerations, the Hamiltonian of the whole system in the
interaction picture and after the rotating wave approximation is given by
(assuming $\hbar =1$)
\begin{equation}
H=\left( g_{1}e^{i\Delta t}\hat{a}_{1}\sigma _{fg}^{+}+g_{2}e^{i\Delta t}%
\hat{a}_{2}\sigma _{fe}^{+}\right) +\left( g_{3}e^{i\Delta ^{\prime }t}\hat{a%
}_{3}\sigma _{fg}^{+}+g_{4}e^{i\Delta ^{\prime }t}\hat{a}_{4}\sigma
_{fe}^{+}\right) +\Omega \sigma _{eg}^{+}+\text{H.c.,}
\end{equation}%
where $\sigma _{fg}^{+}=\left\vert f\right\rangle \left\langle g\right\vert $%
, $\sigma _{fe}^{+}=\left\vert f\right\rangle \left\langle e\right\vert ,$ $%
\sigma _{eg}^{+}=\left\vert e\right\rangle \left\langle g\right\vert ,$ $%
\Omega $ is the Rabi frequency of the pulse, and $\hat{a}_{j}$ is the photon
annihilation operator of cavity $j$ ($j=1,2,3,4$).

In the large-detuning case of $\Delta \gg g_{1},g_{2}$ and $\left\vert
\Delta ^{\prime }\right\vert $ $\gg g_{3},g_{4},$ the intermediate level $%
\left\vert f\right\rangle $ can be adiabatically eliminated. As a result,
the Raman couplings between the states $\left\vert g\right\rangle $ and $%
\left\vert e\right\rangle $ are induced by the cavity pairs $\left(
1,2\right) $ and $\left( 3,4\right) $. When $\frac{\left\vert \Delta -\Delta
^{\prime }\right\vert }{\left\vert \Delta ^{-1}+\Delta ^{\prime
-1}\right\vert }\gg g_{1}g_{4},g_{2}g_{3},$the Raman couplings between the
states $\left\vert g\right\rangle $ and $\left\vert e\right\rangle ,$ caused
by the cavity pairs $\left( 1,4\right) $ and $\left( 2,3\right) ,$ are
suppressed because the corresponding effective coupling strengths are much
smaller than the detunings of these Raman transitions. We assume $\Delta
,\left\vert \Delta ^{\prime }\right\vert \gg \Omega ,$ for which the effect
of the pulse on the Raman couplings is negligible. Thus, we can obtain the
following effective Hamiltonian~\cite{s82}
\begin{eqnarray}
H_{\mathrm{eff}} &=&-2\lambda _{1}\hat{a}_{1}^{+}\hat{a}_{1}\sigma
_{gg}-2\lambda _{2}\hat{a}_{2}^{+}\hat{a}_{2}\sigma _{ee}  \notag
\label{eq2} \\
&&-2\lambda _{3}\hat{a}_{3}^{+}\hat{a}_{3}\sigma _{gg}-2\lambda _{4}\hat{a}%
_{4}^{+}\hat{a}_{4}\sigma _{ee}  \notag \\
&&-2\lambda (\hat{a}_{1}\hat{a}_{2}^{+}\sigma _{eg}^{+}+\hat{a}_{1}^{+}\hat{a%
}_{2}\sigma _{eg}^{-})  \notag \\
&&-2\lambda ^{\prime }(\hat{a}_{3}\hat{a}_{4}^{+}\sigma _{eg}^{+}+\hat{a}%
_{3}^{+}\hat{a}_{4}\sigma _{eg}^{-})  \notag \\
&&+\Omega \sigma _{x},
\end{eqnarray}%
where $\sigma _{eg}^{-}=\left\vert g\right\rangle \left\langle e\right\vert
,$ $\sigma _{gg}=\left\vert g\right\rangle \left\langle g\right\vert
, $~$\sigma _{ee}=\left\vert e\right\rangle \left\langle e\right\vert
, $  $\sigma _{x}=\sigma _{eg}^{+}+\sigma _{eg}^{-},$\ $\lambda
_{1}=g_{1}^{2}/\left( 2\Delta \right) ,$ $\lambda _{2}=g_{2}^{2}/\left(
2\Delta \right) ,$ $\lambda _{3}=g_{3}^{2}/\left( 2\Delta ^{\prime }\right)
, $ $\lambda _{4}=g_{4}^{2}/\left( 2\Delta ^{\prime }\right) ,$ $\lambda
=g_{1}g_{2}/\left( 2\Delta \right) ,$ and $\lambda ^{\prime
}=g_{3}g_{4}/2\Delta ^{\prime }.$ Here, the terms in the first line are
ac-Stark shifts of the level $\left\vert g\right\rangle $\ ($\left\vert
e\right\rangle $) induced by the cavity $1$\ ($2$), the terms in the second
line are ac-Stark shifts of the level $\left\vert g\right\rangle $\ ($%
\left\vert e\right\rangle $) induced by the cavity $3$\ ($4$), the terms in
the third line represent the Raman coupling induced by the cavity pair $%
(1,2) $, while the terms in the fourth line represent the Raman coupling
induced by the cavity pair $(3,4).$

With definition of $|\pm \rangle =(|g\rangle \pm |e\rangle )/\sqrt{2},$ the
operators of the coupler qutrit in Eq. (2) can be expressed as $\sigma
_{gg}=\left( I+\widetilde{\sigma }^{+}+\widetilde{\sigma }^{-}\right) /2,$ $%
\sigma _{ee}=\left( I-\widetilde{\sigma }^{+}-\widetilde{\sigma }^{-}\right)
/2,$ $\sigma _{eg}^{+}=\left( \widetilde{\sigma }_{z}+\widetilde{\sigma }%
^{+}-\widetilde{\sigma }^{-}\right) /2,$ $\sigma _{eg}^{-}=\left( \widetilde{%
\sigma }_{z}-\widetilde{\sigma }^{+}+\widetilde{\sigma }^{-}\right) /2$, and
$\sigma _{x}=\widetilde{\sigma }_{z}$, where $\widetilde{\sigma }%
_{z}=|+\rangle \langle +|-|-\rangle \langle -|,$ $\widetilde{\sigma }%
^{+}=|+\rangle \langle -|,$ and $\widetilde{\sigma }^{-}=|-\rangle \langle
+| $. Using these expressions, one can rewrite Eq. (2), which will contain
the terms $e^{-i2\Omega t}$ and $e^{i2\Omega t}.$ In the strong driving
regime $2\Omega \gg \lambda _{1},\lambda _{2},|\lambda _{3}|,|\lambda
_{4}|,\lambda ,|\lambda ^{\prime }|,$these terms oscillate with high
frequencies and can be discarded after applying a rotating-wave
approximation. Thus, it is easy to find that the Hamiltonian (2) becomes
\begin{eqnarray}
\widetilde{H}_{\mathrm{eff}} &=&-\left( \lambda _{1}\hat{a}_{1}^{+}\hat{a}%
_{1}+\lambda _{2}\hat{a}_{2}^{+}\hat{a}_{2}+\lambda _{3}\hat{a}_{3}^{+}\hat{a%
}_{3}+\lambda _{4}\hat{a}_{4}^{+}\hat{a}_{4}\right) +\Omega \widetilde{%
\sigma }_{z}  \notag \\
&&-\lambda \left( \hat{a}_{1}\hat{a}_{2}^{+}+\hat{a}_{1}^{+}\hat{a}%
_{2}\right) \widetilde{\sigma }_{z}-\lambda ^{\prime }\left( \hat{a}_{3}\hat{%
a}_{4}^{+}+\hat{a}_{3}^{+}\hat{a}_{4}\right) \widetilde{\sigma }_{z}.
\end{eqnarray}

Performing a unitary transformation $e^{iH_{0}t}$, with $H_{0}=-\left(
\lambda _{1}\hat{a}_{1}^{+}\hat{a}_{1}+\lambda _{2}\hat{a}_{2}^{+}\hat{a}%
_{2}+\lambda _{3}\hat{a}_{3}^{+}\hat{a}_{3}+\lambda _{4}\hat{a}_{4}^{+}\hat{a%
}_{4}\right) +\Omega \widetilde{\sigma }_{z},$ we obtain
\begin{eqnarray}
H_{e} &=&e^{iH_{0}t}\left( \widetilde{H}_{\mathrm{eff}}-H_{0}\right)
e^{-iH_{0}t}  \notag \\
\ \ &=&-\lambda \left( \hat{a}_{1}\hat{a}_{2}^{+}+\hat{a}_{1}^{+}\hat{a}%
_{2}\right) \widetilde{\sigma }_{z}+\lambda \left( \hat{a}_{3}\hat{a}%
_{4}^{+}+\hat{a}_{3}^{+}\hat{a}_{4}\right) \widetilde{\sigma }_{z},
\end{eqnarray}%
where we have set
\begin{eqnarray}
\lambda _{1} &=&\lambda _{2},\text{ }\lambda _{3}=\lambda _{4},\text{ } \\
\lambda &=&-\lambda ^{\prime }.
\end{eqnarray}

The qutrit is in the state $\left\vert +\right\rangle ,$ which can be easily
prepared by applying a $\pi $-pulse resonant with the $\left\vert
g\right\rangle \leftrightarrow \left\vert e\right\rangle $ transition of the
qutrit initially in the ground state $\left\vert g\right\rangle $. Note that
the qutrit remains in the state $\left\vert +\right\rangle $ because this
state is not affected by the Hamiltonian~(4). Hence, the qutrit part can be
ignored and the effective Hamiltonian~(4) further reduces to
\begin{equation}
H_{e}=H_{e1}+H_{e2},
\end{equation}%
with
\begin{eqnarray}
H_{e1} &=&-\lambda \left( \hat{a}_{1}\hat{a}_{2}^{+}+\hat{a}_{1}^{+}\hat{a}%
_{2}\right) , \\
H_{e2} &=&\lambda \left( \hat{a}_{3}\hat{a}_{4}^{+}+\hat{a}_{3}^{+}\hat{a}%
_{4}\right) .
\end{eqnarray}%
This Hamiltonian~(7) describes the qutrit-mediated effective interaction
between the cavity pair ($1,2$) and the qutrit-mediated effective
interaction between the cavity pair ($3,4$), which will be used below to
transfer quantum entangled states of two cqubits from the two cavities ($1,2$%
) to the other two cavities ($3,4$).

Initially, cavities $1$ and $3$ store the maximally-entangled state $\left(
|cat\rangle _{1}|cat\rangle _{3}+|\overline{cat}\rangle _{1}|\overline{cat}%
\rangle _{3}\right) /\sqrt{2}$ of two cqubits while cavities $2$ and $4$ are
initially in the vacuum state $|0\rangle _{2}|0\rangle _{4}.$ Here, the two
cat states are given by $|cat\rangle =N_{\alpha }^{+}(|\alpha \rangle
+|-\alpha \rangle )$ and $|\overline{cat}\rangle =N_{\alpha }^{-}(|\alpha
\rangle -|-\alpha \rangle ),$ with the normalization coefficients $N_{\alpha
}^{\pm }$. In terms of $|\alpha \rangle =e^{-|\alpha
|^{2}/2}\sum\limits_{n=0}^{\infty }\frac{\alpha ^{n}}{\sqrt{n!}}|n\rangle $
and $|-\alpha \rangle =e^{-|\alpha |^{2}/2}\sum\limits_{n=0}^{\infty }\frac{%
(-\alpha )^{n}}{\sqrt{n!}}|n\rangle ,$ we have
\begin{equation}
|cat\rangle =\sum\limits_{m=0}^{\infty }C_{2m}|2m\rangle ,\text{ }|\overline{%
cat}\rangle =\sum\limits_{n=0}^{\infty }C_{2n+1}|2n+1\rangle ,
\end{equation}%
where $n$ and $m$ are non-negative integers, $C_{2m}=2N_{\alpha
}^{+}e^{-|\alpha |^{2}/2}\alpha ^{2m}/\sqrt{(2m)!},$ and $C_{2n+1}=$ $%
2N_{\alpha }^{-}e^{-|\alpha |^{2}/2}\alpha ^{2n+1}/\sqrt{(2n+1)!}$. From
Eq.~(10) one can see that the cat state $|cat\rangle $ is orthogonal to the
cat state $|\overline{cat}\rangle $, independent of $\alpha $ (except for $%
\alpha =0$). The two cat states considered here are called even and odd
coherent states in quantum optics.

The transfer of the two-cqubit entangled state from two cavities 1 and 3 to
the other two cavities 2 and 4 is described by
\begin{equation}
\frac{1}{\sqrt{2}}\left( |cat\rangle _{1}|cat\rangle _{3}+|\overline{cat}%
\rangle _{1}|\overline{cat}\rangle _{3}\right) |0\rangle _{2}|0\rangle
_{4}\rightarrow |0\rangle _{1}|0\rangle _{3}\frac{1}{\sqrt{2}}\left(
|cat\rangle _{2}|cat\rangle _{4}+|\overline{cat}\rangle _{2}|\overline{cat}%
\rangle _{4}\right) .
\end{equation}%
In the following, we will show how this entangled state transfer can be
achieved.

According to Eq.~(10) and because of $|2m\rangle _{j}=\left( \hat{a}%
_{j}^{+}\right) ^{2m}|0\rangle _{j}/\sqrt{\left( 2m\right) !}$ and $%
|2n+1\rangle _{j}=\left( \hat{a}_{j}^{+}\right) ^{2n+1}|0\rangle _{j}/\sqrt{%
\left( 2n+1\right) !},$ the two cat states $|cat\rangle _{j}$ and $|%
\overline{cat}\rangle _{j}$ of cavity $j$ ($j=1,2,3,4$) can be expressed as
\begin{eqnarray}
|cat\rangle _{j} &=&\sum\limits_{m=0}^{\infty }C_{2m}^{\prime }\left( \hat{a}%
_{j}^{+}\right) ^{2m}|0\rangle _{j},  \notag  \label{eq12} \\
|\overline{cat}\rangle _{j} &=&\sum\limits_{n=0}^{\infty }C_{2n+1}^{\prime
}\left( \hat{a}_{j}^{+}\right) ^{2n+1}|0\rangle _{j},
\end{eqnarray}%
where $C_{2m}^{\prime }=C_{2m}/\sqrt{\left( 2m\right) !}$ and $%
C_{2n+1}^{\prime }=C_{2n+1}/\sqrt{\left( 2n+1\right) !}.$

Under the effective Hamiltonian $H_{e}$ of Eq. ~(7) and because of $\left[
H_{e1},H_{e2}\right] =0,$ we have the following state evolution%
\begin{eqnarray}
&&e^{-iH_{e}t}|cat\rangle _{1}|cat\rangle _{3}|0\rangle _{2}|0\rangle _{4}
\notag  \label{eq13} \\
&=&e^{-iH_{e1}t}|cat\rangle _{1}|0\rangle _{2}e^{-iH_{e2}t}|cat\rangle
_{3}|0\rangle _{4}  \notag \\
&=&\sum\limits_{m=0}^{\infty }C_{2m}^{\prime }e^{-iH_{e1}t}\left( \hat{a}%
_{1}^{+}\right) ^{2m}e^{iH_{e1}t}\otimes e^{-iH_{e1}t}|0\rangle
_{1}|0\rangle _{2}  \notag \\
&&\otimes \sum\limits_{m=0}^{\infty }C_{2m}^{\prime }e^{-iH_{e2}t}\left(
\hat{a}_{3}^{+}\right) ^{2m}e^{iH_{e2}t}\otimes e^{-iH_{e2}t}|0\rangle
_{3}|0\rangle _{4}  \notag \\
&=&\sum\limits_{m=0}^{\infty }C_{2m}^{\prime }\left( e^{-iH_{e1}t}\hat{a}%
_{1}^{+}e^{iH_{e1}t}\right) ^{2m}|0\rangle _{1}|0\rangle _{2}  \notag \\
&&\otimes \sum\limits_{m=0}^{\infty }C_{2m}^{\prime }\left( e^{-iH_{e2}t}%
\hat{a}_{3}^{+}e^{iH_{e2}t}\right) ^{2m}|0\rangle _{3}|0\rangle _{4},
\end{eqnarray}%
and%
\begin{eqnarray}
&&e^{-iH_{e}t}|\overline{cat}\rangle _{1}|\overline{cat}\rangle
_{2}|0\rangle _{3}|0\rangle _{4}  \notag  \label{eq14} \\
&=&e^{-iH_{e1}t}|\overline{cat}\rangle _{1}|0\rangle _{2}e^{-iH_{e2}t}|%
\overline{cat}\rangle _{3}|0\rangle _{4}  \notag \\
&=&\sum\limits_{n=0}^{\infty }C_{2n+1}^{\prime }e^{-iH_{e1}t}\left( \hat{a}%
_{1}^{+}\right) ^{2n+1}e^{iH_{e1}t}\otimes e^{-iH_{e1}t}|0\rangle
_{1}|0\rangle _{2}  \notag \\
&&\otimes \sum\limits_{n=0}^{\infty }C_{2n+1}^{\prime }e^{-iH_{e2}t}\left(
\hat{a}_{3}^{+}\right) ^{2n+1}e^{iH_{e2}t}\otimes e^{-iH_{e2}t}|0\rangle
_{3}|0\rangle _{4}  \notag \\
&=&\sum\limits_{n=0}^{\infty }C_{2n+1}^{\prime }\left( e^{-iH_{e1}t}\hat{a}%
_{1}^{+}e^{iH_{e1}t}\right) ^{2n+1}|0\rangle _{1}|0\rangle _{2}  \notag \\
&&\otimes \sum\limits_{n=0}^{\infty }C_{2n+1}^{\prime }\left( e^{-iH_{e2}t}%
\hat{a}_{3}^{+}e^{iH_{e2}t}\right) ^{2n+1}|0\rangle _{3}|0\rangle _{4},
\end{eqnarray}%
where we have used $e^{-iH_{e1}t}|0\rangle _{1}|0\rangle _{2}=|0\rangle
_{1}|0\rangle _{2}$ and $e^{-iH_{e2}t}|0\rangle _{3}|0\rangle _{4}=|0\rangle
_{3}|0\rangle _{4}.$

Note that $e^{-iH_{e1}t}\hat{a}_{1}^{\dagger }e^{iH_{e1}t}=\cos (\lambda t)%
\hat{a}_{1}^{\dagger }+i\sin (\lambda t)\hat{a}_{2}^{\dagger }$ and $%
e^{-iH_{e2}t}\hat{a}_{3}^{\dagger }e^{iH_{e2}t}=\cos (\lambda t)\hat{a}%
_{3}^{\dagger }-i\sin (\lambda t)\hat{a}_{4}^{\dagger }.$ For $\lambda t=\pi
/2,$ we have $e^{-iH_{e1}t}\hat{a}_{1}^{\dagger }e^{iH_{e1}t}=i\hat{a}%
_{2}^{\dagger }$ and $e^{-iH_{e2}t}\hat{a}_{3}^{\dagger }e^{iH_{e2}t}=-i\hat{%
a}_{4}^{\dagger }.$ Thus, for $t=T=\pi /(2\lambda ),$ we have from Eqs.~(13)
and~(14)
\begin{eqnarray}
&&e^{-iH_{e}t}|cat\rangle _{1}|cat\rangle _{3}|0\rangle _{2}|0\rangle _{4}
\notag  \label{eq15} \\
&=&|0\rangle _{1}\sum\limits_{m=0}^{\infty }C_{2m}^{\prime }e^{im\pi }\left(
\hat{a}_{2}^{\dagger }\right) ^{2m}|0\rangle _{2}\otimes |0\rangle
_{3}\sum\limits_{m=0}^{\infty }C_{2m}^{\prime }e^{-im\pi }\left( \hat{a}%
_{4}^{\dagger }\right) ^{2m}|0\rangle _{4}  \notag \\
&=&|0\rangle _{1}\sum\limits_{m=0}^{\infty }e^{im\pi }C_{2m}|2m\rangle
_{2}\otimes |0\rangle _{3}\sum\limits_{m=0}^{\infty }e^{-im\pi
}C_{2m}|2m\rangle _{4},
\end{eqnarray}%
and
\begin{eqnarray}
&&e^{-iH_{e}t}|\overline{cat}\rangle _{1}|\overline{cat}\rangle
_{2}|0\rangle _{3}|0\rangle _{4}  \notag  \label{eq16} \\
&=&|0\rangle _{1}\sum\limits_{n=0}^{\infty }C_{2n+1}^{\prime }e^{i\left(
2n+1\right) \pi /2}\left( \hat{a}_{2}^{\dagger }\right) ^{2n+1}|0\rangle
_{2}\otimes |0\rangle _{3}\sum\limits_{n=0}^{\infty }C_{2n+1}^{\prime
}e^{-i\left( 2n+1\right) \pi /2}\left( \hat{a}_{4}^{\dagger }\right)
^{2n+1}|0\rangle _{4}  \notag \\
&=&|0\rangle _{1}\sum\limits_{n=0}^{\infty }e^{i\left( 2n+1\right) \pi
/2}C_{2n+1}|2n+1\rangle _{2}\otimes |0\rangle _{3}\sum\limits_{n=0}^{\infty
}e^{-i\left( 2n+1\right) \pi /2}C_{2n+1}|2n+1\rangle _{4},
\end{eqnarray}%
where we have used $|2m\rangle _{j}=\left( \hat{a}_{j}^{+}\right)
^{2m}|0\rangle _{j}/\sqrt{\left( 2m\right) !},$ $|2n+1\rangle _{j}=\left(
\hat{a}_{j}^{+}\right) ^{2n+1}|0\rangle _{j}/\sqrt{\left( 2n+1\right) !}$ ($%
j=2,4$)$,$ and the definitions of $C_{2m}^{\prime }$ and $C_{2n+1}^{\prime }$
given above.

After returning to the original interaction picture, the time evolution for
the initial state of the whole system is given by
\begin{eqnarray}
&&e^{-iH_{0}\tau }e^{-iH_{e}\tau }\frac{1}{\sqrt{2}}\left( |cat\rangle
_{1}|cat\rangle _{3}+|\overline{cat}\rangle _{1}|\overline{cat}\rangle
_{3}\right) |0\rangle _{2}|0\rangle _{4}\left\vert +\right\rangle  \notag
\label{eq17} \\
&=&e^{-iH_{0}\tau }\frac{1}{\sqrt{2}}\left( |0\rangle
_{1}\sum\limits_{m=0}^{\infty }e^{im\pi }C_{2m}|2m\rangle _{2}\otimes
|0\rangle _{3}\sum\limits_{m=0}^{\infty }e^{-im\pi }C_{2m}|2m\rangle
_{4}\right.  \notag \\
&&\left. +|0\rangle _{1}\sum\limits_{n=0}^{\infty }e^{i\left( 2n+1\right)
\pi /2}C_{2n+1}|2n+1\rangle _{2}\otimes |0\rangle
_{3}\sum\limits_{n=0}^{\infty }e^{-i\left( 2n+1\right) \pi
/2}C_{2n+1}|2n+1\rangle _{4}\right) \left\vert +\right\rangle  \notag \\
&=&e^{-i\phi _{0}}|0\rangle _{1}|0\rangle _{3}\frac{1}{\sqrt{2}}\left[
\sum\limits_{m=0}^{\infty }e^{i2\eta m\pi }C_{2m}|2m\rangle _{2}\otimes
\sum\limits_{m=0}^{\infty }e^{i2\eta ^{\prime }m\pi }C_{2m}|2m\rangle
_{4}\right.  \notag \\
&&\left. +\sum\limits_{n=0}^{\infty }e^{i\eta \left( 2n+1\right) \pi
}C_{2n+1}|2n+1\rangle _{2}\otimes \sum\limits_{n=0}^{\infty }e^{i\eta
^{\prime }\left( 2n+1\right) \pi }C_{2n+1}|2n+1\rangle _{4}\right]
\left\vert +\right\rangle ,
\end{eqnarray}%
where from line 1 to lines 2 and 3 we have used the results given in
Eqs.~(15) and~(16). Here, $\phi _{0}=\Omega \pi /(2\lambda ),$ $\eta
=\lambda _{2}/(2\lambda )+1/2,$ and $\eta ^{\prime }=\lambda _{4}/(2\lambda
)-1/2.$ We set
\begin{eqnarray}
\lambda _{2} &=&\lambda ,\text{ } \\
\lambda _{4} &=&-\lambda ,
\end{eqnarray}%
which leads to $\eta =1$ and $\eta ^{\prime }=-1.$ For $\eta =1$ and $\eta
^{\prime }=-1,$ we have $\exp (i2\eta m\pi )=\exp (i2\eta ^{\prime }m\pi )=1$
and $\exp \left[ i\eta \left( 2n+1\right) \pi \right] =\exp \left[ i\eta
^{\prime }\left( 2n+1\right) \pi \right] =-1.$ Thus, the state of the four
cavities, given in Eq. (17), becomes
\begin{equation}
|0\rangle _{1}|0\rangle _{3}\frac{1}{\sqrt{2}}\left(
\sum\limits_{m=0}^{\infty }C_{2m}|2m\rangle _{2}\otimes
\sum\limits_{m=0}^{\infty }C_{2m}|2m\rangle _{4}+\sum\limits_{n=0}^{\infty
}C_{2n+1}|2n+1\rangle _{2}\otimes \sum\limits_{n=0}^{\infty
}C_{2n+1}|2n+1\rangle _{4}\right) ,
\end{equation}%
where the common phase factor $e^{-i\phi _{0}}$ has been omitted. According
to Eq.~(12), the state~(20) can be written as
\begin{equation}
|0\rangle _{1}|0\rangle _{3}\frac{1}{\sqrt{2}}\left( |cat\rangle
_{2}|cat\rangle _{4}+|\overline{cat}\rangle _{2}|\overline{cat}\rangle
_{4}\right) ,
\end{equation}%
which shows that the quantum entangled state of two cqubits, originally
stored in the two cavities 1 and 3, has been transferred onto the other two
cavities 2 and 4. In order to maintain the transferred state, the level spacings of the
flux qutrit need to be rapidly adjusted~\cite{s80} so that the qutrit
is decoupled from four cavities after the desired state transfer is completed.
Alternatively, to have the cavities coupled or decoupled from the qutrit, one can also
tune the frequencies of cavities~\cite{s81}.

The result (21) was derived under the conditions (5), (6), (18) and (19)
given above. The conditions (5) and (18) can be satisfied by choosing $%
g_{1}=g_{2}$ and $g_{3}=g_{4}.$ This requirement for the coupling constants
can be achieved for either 3D cavities or 1D cavities. For 3D cavities, $%
g_{j}$ can be adjusted by a prior design of the sample with a suitable loop
area of the qutrit that falls in cavity $j$ $(j=1,2,3,4)$. In addition, for
1D cavities, $g_{j}$ can be adjusted by a prior design of the sample with an
appropriate capacitance $C_{j}$ between the qutrit and cavity $j$ $%
(j=1,2,3,4)$.

One can check that both conditions (6) and (19) turn out into
\begin{equation}
g_{1}g_{2}/\Delta =-g_{3}g_{4}/\Delta ^{\prime },
\end{equation}%
i.e.,
\begin{equation}
g_{1}g_{2}/\left( \omega _{fg}-\omega _{1}\right) =g_{1}g_{2}/\left( \omega
_{fe}-\omega _{2}\right) =g_{3}g_{4}/\left( \omega _{3}-\omega _{fg}\right)
=g_{3}g_{4}/\left( \omega _{4}-\omega _{fe}\right) ,
\end{equation}%
which can be met by adjusting the cavity frequencies, the qutrit level
spacings, or the coupling constants.

From the above description, one can see that the coupler qutrit remains in
the state $\left\vert +\right\rangle $ during the state transfer. In other
words, the level $\left\vert f\right\rangle $ of the qutrit is not excited
and thus decoherence from this higher energy level is greatly suppressed.
Eq.~(\ref{eq17}) shows that the state transfer is performed by applying a
unitary operator $U=e^{-iH_{0}\tau }e^{-iH_{e}\tau }$ on the initial state
of the whole system. As mentioned above, the transformation $e^{-iH_{0}\tau
\text{ \ }}$here is only used in order to return to the original interaction
picture. Thus, it can be concluded that the state transfer is realized with
a single-step operation, described by $U.$

\begin{figure}[tbp]
\includegraphics[bb=48 180 458 530, width=10.5 cm, clip]{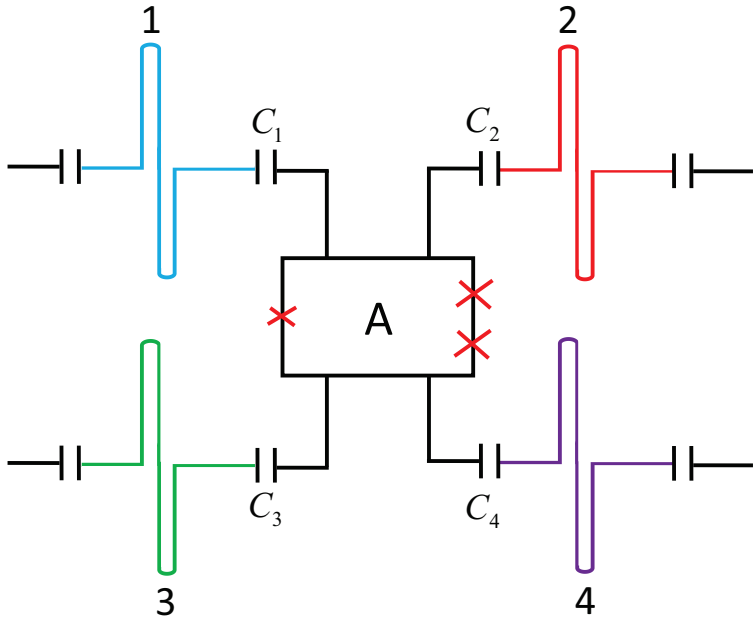} \vspace*{%
-0.08in}
\caption{(Color online) Setup of four transmission line resonators (TLRs)
capacitively coupled to a superconducting flux qutrit (a square $A$ at the
center). The flux qutrit consists of three Josephson junctions and a
superconducting loop.}
\label{fig2}
\end{figure}

\section{Possible experimental implementation}

In above, we have considered a general type of cavity, either 3D cavity or
1D cavity. In this section, we consider a setup of four transmission line
resonators (TLRs) capacitively coupled to a superconducting flux qutrit
(Fig. 2). Each TLR here is a 1D microwave cavity.
For a flux device (e.g., C-shunted flux qubit \cite{s83,s84,s85}), the level spacings can be designed to have a sufficiently large
anharmonicity and the transition between non-adjacent levels is allowed.
Accordingly, our proposal uses the flux qutrit so that the resonator coupling with the flux qutrit's $|g\rangle\leftrightarrow|f\rangle$ transition is available.
In the following, we will
give a discussion on the experimental feasibility of transferring quantum
entangled states of two cqubits between the two TLRs.

By taking the unwanted interaction into account, the Hamiltonian~(1) is
modified as $H^{\prime }=H+\delta \!H_{1}+\delta \!H_{2}$. Here, $\delta
\!H_{1}$ describes the unwanted inter-cavity crosstalk, with the form of
\begin{equation}
\delta \!H_{1}=\sum\limits_{l=j+1}^{4}\sum\limits_{j=1}^{4}g_{jl}e^{i\Delta
_{jl}t}\hat{a}_{j}\hat{a}_{l}^{+}+\text{H.c.},
\end{equation}%
where $g_{jl}$ and $\Delta _{jl}=\omega _{l}-\omega _{j}$ are, respectively,
the coupling strength and the frequency detuning of the two cavities $j$ and
$l$ ($jl=12,13,14,23,24,34$). In addition, $\delta \!H_{2}$ describes the
unwanted $\left\vert e\right\rangle \leftrightarrow \left\vert
f\right\rangle $ transition induced by the pulse, which is given by
\begin{equation}
\delta \!H_{2}=\Omega _{fe}e^{i\Delta_p t}S_{fe}^{+}+\text{H.c.},
\end{equation}%
where $\Delta_p =\omega _{fe}-\omega _{eg},$ and $\Omega _{fe}$ is the Rabi
frequency of the pulse, associated with the $\left\vert e\right\rangle
\leftrightarrow \left\vert f\right\rangle $ transition.

Because of $\omega _{eg}\ll \omega _{fg}$ [Fig.~\ref{fig1}(b)], the $%
\left\vert g\right\rangle \leftrightarrow \left\vert f\right\rangle $
transition induced by the pulse is negligible. For a superconducting flux
device, the level spacings can be designed to have a sufficiently large
anharmonicity, such that the cavity-induced coherent transitions between any
other irrelevant levels are negligibly small. Hence, the effects of the
cavity-induced unwanted transitions as well as the pulse-induced $\left\vert
g\right\rangle \leftrightarrow \left\vert f\right\rangle $ transition on the
state transfer performance are negligible and thus not considered in our
numerical simulations for simplicity.

After considering the qutrit dephasing and energy relaxation and the cavity
dissipation, the system dynamics under Markovian approximation is governed
by the master equation
\begin{eqnarray}  \label{eq23}
\frac{d\rho }{dt} &=&-i\left[ H^{\prime },\rho \right] +\sum%
\limits_{j=1}^{4}\kappa _{j}\mathcal{L}\left[ \hat{a}_{j}\right]  \notag \\
&&+\gamma _{fe}\mathcal{L}\left[ \sigma _{fe}^{-}\right] +\gamma _{fg}%
\mathcal{L}\left[ \sigma _{fg}^{-}\right] +\gamma _{eg}\mathcal{L}\left[
\sigma _{eg}^{-}\right]  \notag \\
&&+\sum\limits_{l=e,f}\gamma _{\varphi ,l}\left( \sigma _{ll}\rho \sigma
_{ll}-\sigma _{ll}\rho /2-\rho \sigma _{ll}/2\right) .
\end{eqnarray}%
Here, $\mathcal{L}\left[ \Lambda \right] =\Lambda \rho \Lambda ^{+}-\Lambda
^{+}\Lambda \rho /2-\rho \Lambda ^{+}\Lambda /2$ (with $\Lambda =\hat{a}%
_{j},\sigma _{fe}^{-},\sigma _{fg}^{-},\sigma _{eg}^{-})$,\ and $\sigma
_{ff}=\left\vert f\right\rangle \left\langle f\right\vert $; $\kappa _{j}$
is the decay rate of cavity $j$ ($j=1,2,3,4$);\ $\gamma _{eg}$\ is the
energy relaxation rate for the level $\left\vert e\right\rangle $\ of the
qutrit, associated with the decay path $\left\vert e\right\rangle
\rightarrow \left\vert g\right\rangle $; $\gamma _{fe}$\ ($\gamma _{fg}$) is
the relaxation rate for the level $\left\vert f\right\rangle $ of the
qutrit, related to the decay path $\left\vert f\right\rangle \rightarrow
\left\vert e\right\rangle $ ($\left\vert f\right\rangle \rightarrow
\left\vert g\right\rangle $); $\gamma _{\varphi ,e}$ ($\gamma _{\varphi ,f}$%
) is the dephasing rate of the level $\left\vert e\right\rangle $ ($%
\left\vert f\right\rangle $)\textbf{. }

\begin{figure}[tbp]
\begin{center}
\includegraphics[width=4.5in]{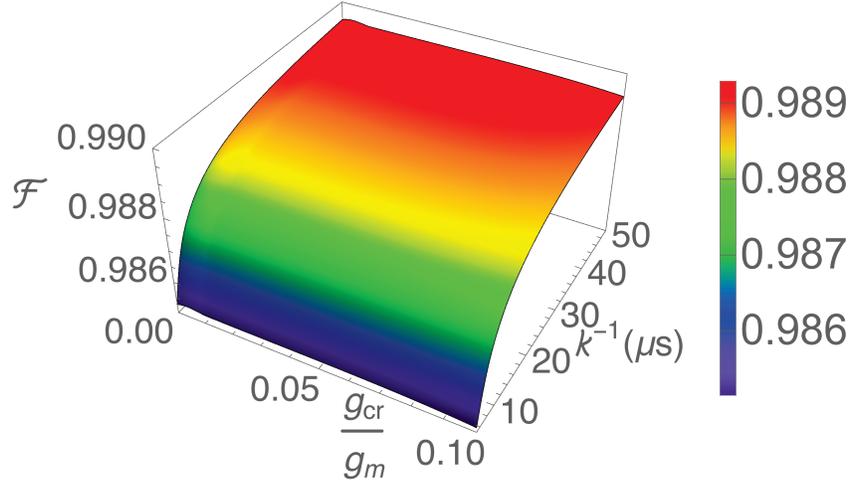} \vspace*{-0.08in}
\end{center}
\caption{(Color online) Fidelity $\mathcal{F}$ versus $\protect\kappa ^{-1}$
and $g_{cr}/g_m$. The parameters used in the numerical simulation are
referred to the text. }
\label{fig3}
\end{figure}

The fidelity of the entangled state transfer is given by $\mathcal{F}=\sqrt{%
\left\langle \psi _{\mathrm{id}}\right\vert \widetilde{\rho }\left\vert \psi
_{\mathrm{id}}\right\rangle },$ where $\left\vert \psi _{\mathrm{id}%
}\right\rangle $ is the ideal output state of the four cavities given in
Eq.~(21), while $\widetilde{\rho }$ is the reduced density operator of the
four cavities after tracing $\rho $ over the degrees of the coupler qutrit,
when the state transfer is carried out in a realistic system (with
dissipation and dephasing considered). Note that our numerical simulations
are performed by choosing the operation time $t=\pi /\left( 2\lambda \right)
$ above.

For a three-level flux qutrit, the transition frequency between two
neighboring levels can be varied from~5 GHz to~20 GHz. As an example, we
consider $\omega _{eg}/2\pi =7.5$ GHz and $\omega _{fg}/2\pi =12.5$ GHz,
resulting in $\Delta _{p}/2\pi =-2.5$ GHz. We choose $\Delta /2\pi =800$
MHz, $g_{1}/2\pi =g_{2}/2\pi =60$ MHz, and $g_{3}/2\pi =g_{4}/2\pi =70$ MHz.
According to Eq.~(22), a simple calculation gives $\Delta ^{\prime }/2\pi
\sim -1.09$ GHz. Note that the coupling constants here are readily
achievable in experiments because a coupling strength $\sim 636$ MHz was
reported for a superconducting flux device coupled to a TLR~\cite{s12}. The
detunings here yields $\omega _{1}/2\pi =11.7$ GHz$,$ $\omega _{2}/2\pi =4.2$
GHz$,$ $\omega _{3}/2\pi =13.59$ GHz$,$ $\omega _{4}/2\pi =6.09$ GHz. Thus,
we have $\Delta _{12}/2\pi =-7.5$ GHz, $\Delta _{13}/2\pi =1.89$ GHz, $%
\Delta _{14}/2\pi =-5.61$ GHz, $\Delta _{23}/2\pi =9.39$ GHz, $\Delta
_{24}/2\pi =1.89$ GHz, and $\Delta _{34}/2\pi =-7.5$ GHz. For simplicity, we
choose $\Omega _{fe}/2\pi =\Omega /2\pi =47$ MHz (available in experiments ~\cite{s86,s87}). Other parameters used in
the numerical simulation are: (i) $\gamma _{\varphi ,e}^{-1}=\gamma
_{\varphi ,f}^{-1}=7$ $\mu $s, $\gamma _{eg}^{-1}=28$ $\mu $s, $\gamma
_{fe}^{-1}=14$ $\mu $s, $\gamma _{fg}^{-1}=21$ $\mu $s (a conservative
consideration, e.g., see Refs.~\cite{s83,s84,s85}); (ii) $\kappa _{1}^{-1}=\kappa
_{2}^{-1}=\kappa _{3}^{-1}=\kappa _{4}^{-1}=\kappa ^{-1}$; and (iii) $\alpha
=1.5.$

We now numerically calculate the fidelity for the entangled state transfer.
For simplicity, we assume that the crosstalk strength for every two cavities
is equal, and thus we set $g_{jl}\equiv g_{cr}$ ($jl=12,13,14,23,24,34)$. To
see how the inter-cavity crosstalk and the cavity decay affect the operation
performance, we plot Fig.~\ref{fig3} to show the fidelity $\mathcal{F}$
versus $\kappa ^{-1}$ and $g_{cr}/g_{m}$. Here, $g_{m}=\max
\{g_{1},g_{2},g_{3},g_{4}\}.$ From Fig.~\ref{fig3}, one can see that the
effect of the inter-cavity crosstalk coupling is very small for a given $%
\kappa ^{-1}.$ One can see that even when $g_{cr}=0.1g_{m}$, a high fidelity
$\sim 98.7\%$ can be reached for $\kappa ^{-1}=10$ $\mu $s. Note that the
crosstalk strength between cavities can be made $0.01g_{m}$ by a prior
design of the sample with appropriate capacitances $C_{1},C_{2},C_{3},C_{4}$~\cite{s54}.

\begin{figure}[tbp]
\begin{center}
\includegraphics[width=3.5in]{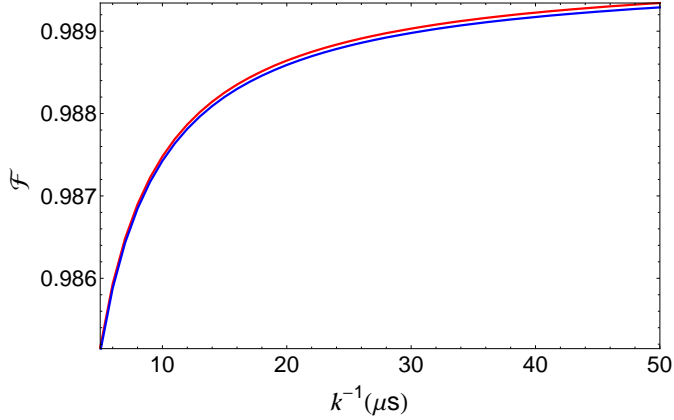} \vspace*{-0.08in}
\end{center}
\caption{(Color online) Fidelity $\mathcal{F}$ versus $\protect\kappa ^{-1}$. The red line correspond to the case without considering
the decay of the second excited state of the qutrit, while the blue line correspond to the case that the decay of the second excited state of the qutrit is taken into account. }
\label{fig4}
\end{figure}
To investigate the effect of the decay of the second excited state (i.e., $\left\vert f\right\rangle$ ) of the qutrit on the fidelity, we numerically calculate the operation fidelity for the entangled state transfer for  (i) $\gamma
_{\varphi ,f}^{-1}=7$ $\mu $s, $\gamma
_{fe}^{-1}=14$ $\mu $s, $\gamma _{fg}^{-1}=21$ $\mu $s and (ii) $\gamma
_{\varphi ,f}^{-1}=\gamma
_{fe}^{-1}=\gamma _{fg}^{-1}=0$ $\mu $s , as the blue and red lines displayed in Fig.~4.
Figure 4 shows the fidelity $\mathcal{F}$ versus $\protect\kappa ^{-1}$, which is
plotted for $g_{cr}=0.01g_{m}$. Other parameters used in the numerical
simulation for Fig. 4 are the same as those used in Fig.~3.
From Fig.~4, one can see that for $\kappa ^{-1}=10$ $\mu $s,
a high fidelity $\sim 98.747\%$ or $\sim 98.748\%$ is achievable for (i) or (ii).
Figure 4 displays the effect of the decay of the second excited state of the qutrit is negligible.

\begin{figure}[tbp]
\begin{center}
\includegraphics[bb=7 84 680 513, width=12.5 cm, clip]{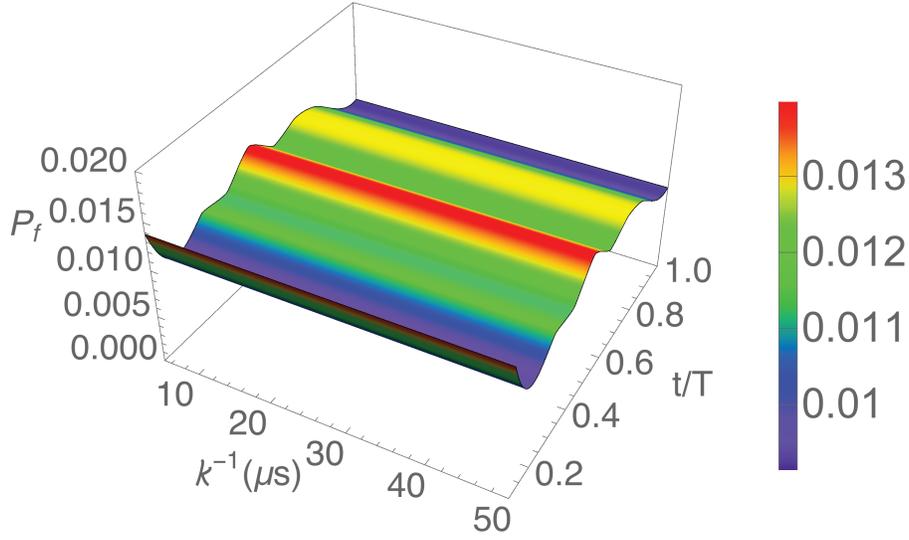} \vspace*{%
-0.08in}
\end{center}
\caption{(Color online) The population $P_f$ of the second excited state of the qutrit versus $t/T$, which is plotted for $g_{cr} =
~0.01g_m$. Other parameters chosen are the
same as those used in Fig.~3. }
\label{fig:5}
\end{figure}
In addition, we also give the numerical results of the population of the second excited state of the qutrit in Fig.~5.
Figure 5 displays the population $P_f$ of the second excited state of the qutrit versus $t/T$ for the entangled state transfer,
which is
plotted for $g_{cr}=0.01g_{m}$.
Here,
$t$ is the state evolution time and $T$ is the entire operation
time required for the state transfer. Other parameters chosen are the
same as those used in Fig.~3.
Figure 5 shows that the population of the second excited state of the qutrit is less than 0.014, implying that the second excited state of the qutrit is almost not populated during the entire operation.
Thus, the decoherence of the qutrit from the second excited state can be efficiently suppressed.

\begin{figure}[tbp]
\begin{center}
\includegraphics[width=3.5in]{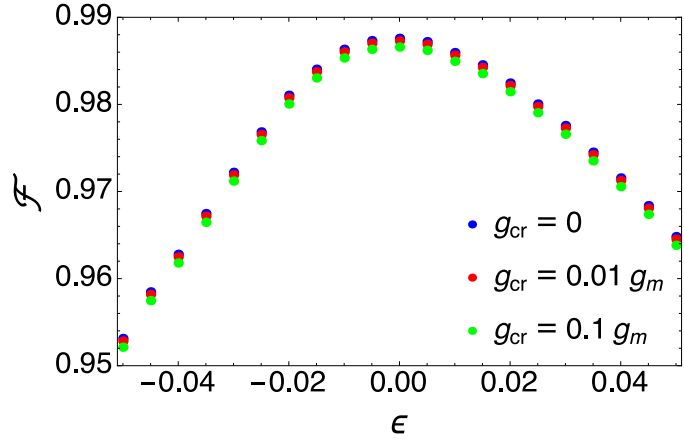} \vspace*{-0.08in}
\end{center}
\caption{(Color online) Fidelity $\mathcal{F}$ versus $\epsilon$. The figure is plotted for $g_{cr} =
0,~0.01g_m,~0.1g_m$ and $\protect\kappa ^{-1} = 10$ $\protect\mu $s. Other
parameters used in the numerical simulation are the same as those used in
Fig.~3.}
\label{fig6}
\end{figure}

In practice, it is an experimental challenge to have $g_{1}=g_{2}$ and $%
g_{3}=g_{4}$. Thus, for the sake of generality, we consider $%
g_{2}=(1+\epsilon )g_{1}$ and $g_{4}=(1+\epsilon )g_{3}$, with the values of
$g_{1}$ and $g_{2}$ given above. We plot Fig.~6 to show how the fidelity $%
\mathcal{F}$ changes with $\epsilon $. Fig.~6 is plotted for $\kappa ^{-1}=10
$ $\mu $s and $g_{cr}=0,0.01g_{m},0.1g_{m}$. Other parameters chosen are the
same as those given in Fig.~3. From Fig.~6, one can see that the fidelity
strongly depends on $\epsilon $, but a high fidelity $\gtrsim 95.3\%$ can still
be available for $-5\%\leq \epsilon \leq 5\%$.

With the parameters chosen above, the operational time is estimated as $0.11$
$\mu $s, much shorter than the decoherence times of the qutrit used in the
numerical simulation and the cavity decay times (5 $\mu $s -- 50 $\mu $s)
considered in Fig.~\ref{fig3}. For the cavity frequencies given above and
for $\kappa _{{}}^{-1}=10$ $\mu $s used in the numerical simulation, the
required quality factors ($Q_j=\kappa_j^{-1}\omega_j$) for the four cavities are $Q_{_{1}}\sim 7.35\times
10^{5},$ $Q_{2}\sim 2.64\times 10^{5}$, $Q_{_{3}}\sim 8.53\times 10^{5},$ $%
Q_{4}\sim 3.82\times 10^{5}.$ The cavity quality factors here are achievable
in experiment because TLRs with a (loaded) quality factor $Q\sim 10^{6}$
have been experimentally demonstrated \cite{s42,s43}. The analysis here
demonstrates that the high-fidelity transfer of quantum entangled states of
two cat-state qubits, from two microwave cavities to the other two microwave
cavities, is feasible within present-day circuit QED techniques.

\section{Conclusion}

We have presented an approach to transfer quantum entangled states of two
cat-state qubits based on circuit QED. As shown above, this proposal has the
advantages stated in the introduction. Our numerical simulations demonstrate
that high-fidelity transfer of quantum entangled states of two cat-state
qubits from two TLRs to the other two TLRs is feasible with current circuit
QED technology. This proposal is quite general and can be applied to
transfer quantum entangled states of two cat-state qubits in a wide range of
physical systems, such as four 1D or 3D (microwave or optical) cavities
coupled to a natural or artificial three-level atom. To the best of our
knowledge, this work is the first to demonstrate the transfer of quantum
entangled states of cat-state qubits, based on circuit or cavity QED. We
hope that this work will stimulate the experimental activity in the near
future.

\section*{Acknowledgments}

This work was partly supported by the Key-Area Research and Development Program of GuangDong Province (2018B030326001), the National Natural Science Foundation of China (NSFC) (11074062, 11374083, 11774076), the NKRDP of China (2016YFA0301802),
and the Jiangxi Natural Science Foundation (20192ACBL20051).

\end{document}